\documentclass[12pt,a4paper]{article}
\usepackage[utf8]{inputenc}
\usepackage[margin=2.2cm]{geometry}
\usepackage{authblk}

\usepackage[breaklinks]{hyperref}
\usepackage{booktabs}
\usepackage{mathptmx}

\usepackage{amsthm}

\newtheorem*{remark*}{Remark}

\usepackage{tikz}
\usetikzlibrary{matrix,arrows,patterns,cd}
\usepackage{leftidx}
\usepackage{amsmath,amssymb,amsfonts}

\usepackage{slashed}
\newcommand{\bbLbrack}{[\kern-0.4em{[}\,}
\newcommand{\bbRbrack}{\,]\kern-0.4em{]}}

\usepackage{bm}
\DeclareMathAlphabet{\mathbfsf}{\encodingdefault}{\sfdefault}{bx}{n}

\usepackage{extarrows}
\usepackage{accents}

\usepackage{mathrsfs,latexsym}
\usepackage[mathscr]{eucal}

\newcommand{\II}{{\boldsymbol{1}}}

\newcommand{\CC}{{\mathbb C}}

\newcommand{\Xc}{{\mathcal X}}

\newcommand{\xb}{{\boldsymbol{x}}}

\DeclareMathOperator{\Tr}{Tr}

\DeclareMathOperator{\Prob}{Prob}

\newcommand{\Af}{{\mathscr A}}

\newcommand{\Bf}{{\mathscr B}}

\newcommand{\Cf}{{\mathscr C}}

\newcommand{\If}{{\mathscr I}}

\newcommand{\Uf}{{\mathscr U}}

\newcommand{\Es}{\mathsf{E}}

\newcommand{\id}{{\rm id}}

\newcommand{\Var}{{\rm Var}}

\makeatletter
\newcommand{\raisemath}[1]{\mathpalette{\raisem@th{#1}}}
\newcommand{\raisem@th}[3]{\raisebox{#1}{$#2#3$}}
\makeatother

\numberwithin{equation}{section}

\begin{document}

\title{Measurement in Quantum Field Theory}
\author[1,2]{Christopher J. Fewster\thanks{\tt chris.fewster@york.ac.uk}}
\affil{Department of Mathematics,
	University of York, Heslington, York YO10 5DD, United Kingdom.}
\affil[2]{York Centre for Quantum Technologies, University of York, Heslington, York YO10 5DD, United Kingdom.}
\author[3]{Rainer Verch\thanks{\tt rainer.verch@uni-leipzig.de}}
\affil[3]{Institute for Theoretical Physics, University of Leipzig, 04009 Leipzig, Germany.}
\date{\today}
\maketitle

\maketitle

\noindent
{\bf Abstract} \ \ The topic of measurement in relativistic quantum field theory is addressed in this article.   Some of the long standing problems of this subject are highlighted, including the incompatibility of an instantaneous ``collapse of the wavefunction'' with relativity of simultaneity, and the difficulty of maintaining causality in the rules for measurement highlighted by ``impossible measurement'' scenarios. Thereafter, the issue is considered from the perspective of mathematical physics. To this end, quantum field theory is described in a model-independent, operator algebraic setting, on generic Lorentzian spacetime manifolds. The process of measurement is modelled by a localized dynamical coupling between a quantum field called the ``system'', and another quantum field, called the ``probe''. The result of the dynamical coupling is a scattering map, whereby measurements carried out on the probe can be interpreted as measurements of induced observables on the system. The localization of the dynamical coupling allows it to derive causal relations for the induced observables. It will be discussed how this approach leads to the concept of selective or non-selective 
system state updates conditioned on the result of probe measurements, which in turn allows it to obtain conditional probabilities for consecutive probe measurements consistent with relativistic causality and general covariance, without the need for a physical collapse of the wavefunction. In particular, the problem of impossible measurements is resolved.
Finally, there is a brief discussion of accelerated detectors and other related work. 

%

\section{Introduction}
 
The topic of {\it measurement in quantum field theory} is quite involved with several different -- albeit sometimes related -- facets. We will address some of them.
\begin{itemize}
\item[{\bf (A)}]  At the onset of the discussion of quantum (or quantized) fields, the question whether, and in which sense,
quantum fields are {\it measurable quantities} was posed in a seminal article by Landau and Peierls~\cite{LandauPeierls}. Their 
argument appears to show that there are fundamental difficulties in measuring quantized electrodynamic fields since they 
must be measured by electrically charged elementary test particles which 
are themselves subject to quantum physics, and on reacting to the quantum field create an electromagnetic field
disturbing the measurement. This line of argument received a  rebuttal by Bohr and Rosenfeld~\cite{BohrRosenf-a}, who pointed
out that a quantized field has (in a more modern way of formally capturing it) a description in terms of 
operator-valued observables at every point in spacetime, and that an operationally meaningful concept of quantum fields
is only achieved by ``macroscopic'' (i.e.\ potentially small, but not strictly pointlike) averaged quantum field measurements.
In a certain sense \cite{BohrRosenf-b}, that insight ultimately led to the description of quantum fields as ``operator valued distributions''~\cite{StreaterWightman}.
See also~\cite{Hartz} for further discussion.

\item[{\bf (B)}] A notorious concept in the description of measurement in non-relativistic quantum mechanics~\cite{vonNeumann:MathFoundQM} is the {\it collapse of the wavefunction}. For quantum field theory, the issue is aggravated by the
need to maintain relativistic covariance and causality; in particular, the need to avoid superluminal transfer of information. We will take up this theme in Sec.\ \ref{Collapse} below. 

\item[{\bf (C)}] A further theme is the {\it process of measurement} of a quantum field by coupling it dynamically to certain ``probe'' systems 
(which may be either classical, quantum mechanical or quantum field theoretical in nature) and investigating the way in which 
measuring of the probe provides information on the quantum fields to be measured. This is the main theme in the present day 
approach to measurement in quantum mechanics \cite{Busch_etal:quantum_measurement}. It has led to multiple insights and helped to remove misunderstandings
in the measurement process in quantum mechanics and its methods are nowadays very widely applied in subjects like quantum information
and quantum control \cite{wiseman_milburn_2009}. Therefore, a promising line of approach towards measurement in quantum field theory is via adopting the methods 
which have proved to be successful for the understanding of measurement in quantum mechanics. We will focus on this approach and we present and discuss some of the progress that has been achieved in this direction. 
It is worth mentioning that the {\it Unruh--DeWitt detector models},
which are used in the description of the {Unruh} effect~\cite{Unruh:1976,CrispinoHiguchiMatsas:2008,EncycMP_QFTCST_Kay:2025}, can also be regarded as probes in the sense just mentioned.
Further discussion to this end will appear in Sec.~\ref{AccDetc}.

\item[{\bf (D)}] Questions about the measurability of quantum fields also arise in the technique of {\it renormalization} in
perturbative approaches to interacting quantum field theories, which are of considerable importance in the application of 
quantum field theory to elementary particle interaction processes. In these approaches, there are systematic algorithms to 
subtract divergent quantities to render the perturbative calculations (e.g.\ scattering cross sections of interaction processes)
well-defined. The subtraction of divergences leaves an ambiguity in the resulting finite values of the renormalized quantities,
typically the masses, charges, field strengths and couplings of the quantum fields representing the elementary particles involved 
in the interaction processes. This ``renormalization ambiguity'' has to be fixed by measurement, thus the question of measurement 
in quantum field theory is of importance also in concrete applications to elementary particle physics
(see, e.g., Sec.\ 5.1 in \cite{PaulRoman_QFT}; see also \cite{Duetsch_book,Rejzner_book} for more modern
approaches to perturbative quantum field theory).
Despite its practical importance, we will not elaborate on this topic. 
\end{itemize}
There are other possible associations with measurement in quantum field theory which we shall not touch upon at all.
For instance, we shall not discuss or analyze specific, experimentally realized quantum field measurements; our discussion is purely theoretical 
and mathematical. Neither will we embark on interpretative, ontological or philosophical aspects of measurements in quantum field theory.
In this spirit, we shall not address matters like ``many world interpretations'' of quantum field theories and their potential implications
for the measurement process. Similarly, we shall not be concerned with problems that may arise in measurement in quantum field theory 
by assuming that space and time were in some sense ``quantized''. 
We always assume quantum field theory to mean ``quantized fields on a classical spacetime'', where 
classical spacetime means the spacetime concept of general relativity. In other words, our theoretical framework is that of quantum field 
theory on curved spacetime (or quantum field theory on Minkowski spacetime as a more specialized variant) \cite{Wald_qft,AdvAQFT,EncycMP_AQFT_BuchholzFredenhagen:2025,EncycMP_AxQFT_Fraser2025,EncycMP_QFTCST_Kay:2025}.

The present article is structured as follows. In the next two sections, we will present two famous problems arising in the measurement process in quantum field theory -- namely, the ``collapse of the wave-function'' and ``measurement and causality''. The latter takes up on a proposed scenario of ``impossible measurements'' in quantum field theory \cite{sorkin1993impossible}. 
In Sec.\ 4, we describe local and covariant measurement schemes and state updates in 
the framework of algebraic quantum field theory according to \cite{FewVer_QFLM:2018}. This approach to measurement in quantum field theory is inspired by the theory of quantum measurement in quantum mechanics \cite{Busch_etal:quantum_measurement}, in which a 
quantum system (the ``system'') is dynamically coupled to another one (the ``probe'').
System and probe interact dynamically and the result of the interaction is a scattering map. With the help of this scattering map, observables of the probe 
give rise to ``induced observables'' of the system, thus allowing one to interpret 
measurements carried out on the probe as measurements (of induced observables) on the
system. While this is to some extent very much related to scattering theory~\cite{EncycMP_AQFT_BuchholzDybalski:2025}, the 
decisive novelty lies in a {\it locality of the coupling} between system and probe. 
Thereby, it is possible to derive causal relations for induced observables. In turn, this leads to a formulation of the measurement of the system by the probe which avoids
the notion of ``collapse of the wavefunction'' entirely, and also allows it to 
show that hypothesized ``impossible measurement'' scenarios do not occur \cite{BostelmannFewsterRuep:2020}. In Sec.\ \ref{AccDetc}, we comment on ``accelerated detectors'' which appear in the context of the Unruh effect. We conclude in Sec.\ \ref{RelAppr} with some remarks on related approaches to measurement in quantum field theory.

\section{``Collapse of the wave-function''}\label{Collapse}

Standard accounts of quantum mechanics teach that the wavefunction
changes instantaneously following a measurement. For a relativistic
theory in Minkowski spacetime, this collapse rule is clearly in conflict with Lorentz covariance; in a curved spacetime it conflicts with the lack of a preferred foliation into spacelike hypersurfaces. 

A variety of views have been taken. Bloch~\cite{Bloch:1967} argued that
wavefunctions were rendered ambiguous in relativistic quantum theory, depending on the frame chosen to analyse the measurement in question. However, this ambiguity would have 
no physical consequences, because consistent predictions of probabilities can be made nonetheless. By contrast, Hellwig and Kraus~\cite{HellwigKraus_prd:1970} posited that the 
state reduction should occur across the backward lightcone of the measurement region, thus providing a manifestly covariant collapse rule that satisfies various consistency conditions. When multiple measurements in disjoint spacetime regions are considered, Hellwig and Kraus divide up the spacetime according to the past lightcones of all the regions concerned and build a `state map' by applying the collapse rule as one crosses any of these surfaces. In some regions, there is potential ambiguity about which collapse rule should be applied first, but it is proved that the same state is obtained whatever the order chosen. 
Hellwig and Kraus note that it is a `pure convention' to leave the state unchanged inside the backward lightcone, observing that an equivalent formalism would have the change within the forward lightcone only. They
are also careful to point out that they discuss only the formal aspects of the rules, without speculating on their physical or philosphical meaning. In particular, they are silent on 
whether one is supposed to believe that some physical entity changes across a particular spacetime surface. 

Aharonov and Albert~\cite{AharonovAlbert:1980,AharonovAlbert:1981} rejected the formalism proposed by Hellwig and Kraus~\cite{HellwigKraus_prd:1970} and the liberal attitude to state ambiguities taken by Bloch~\cite{Bloch:1967}. They consider, for example, a charged particle that is in a superposition of two states, which separately describe the motion of the particle along different worldlines in Minkowski spacetime. A measurement of position causes a reduction to one or other of these worldlines, supposed to occur instantaneously in some inertial frame of reference. Thus the expected current density vector field changes discontinuously from one in which half the charge is concentrated near each worldline, to one in which all the charge is concentrated near one of them. Unsurprisingly, integrals over different Cauchy surfaces can give different values for the expected total charge, conflicting with charge conservation. This happens in particular for any hypersurface that is crossed by one worldline before the instantaneous change, and by the other afterwards.  
Similar problems are found when the collapse is supposed to occur across a backward lightcone. Aharonov and Albert drew the conclusion that no covariant state history can be defined that accounts for experimental results. 

This problem is removed if one relinquishes the idea that 
there is a physical change in the state after measurement, and adopts
the more pragmatic stance that, on the one hand, states are mathematical objects from which probabilities of measurement outcomes may be computed, and on the other
that the foregoing account has ignored what is actually a physical process, namely the interaction between a quantum system and the probe that is used to measure it. (This point has been emphasised in quantum mechanics by many authors, and in quantum field theory e.g., by~\cite{DoplicherQFM}. It is also important in the response of 
Bohr and Rosenfeld to Landau and Peierls mentioned in our opening point {\bf (A)}.)
In a full treatment, to be described later, one models this interaction; if desired, one can trace out the probe using the results of the measurement (to the extent they are known), which amounts to an update of the system state. In this way one obtains an effective description at the level of the system on its own, in which the state is updated in the light of information gained during the measurement. This is an exercise in book-keeping -- in fact, a way of arranging the calculation of conditional probabilities -- that provides an effective description of a physical process.
 A viewpoint of this type has
been emphasised particularly by Peres and Terno~\cite{PeresTerno:2004} (see also \cite{AnastopoulosHuSavvidou:2022}) and we will return to it later. In the situation described above, one makes exactly the same change of states as described by Aharonov and Albert, with the difference that the `before' and `after' states are each defined globally on spacetime and have perfectly smooth and conserved expected current density vector fields; one does not attempt to weld them together across a specific spacetime hypersurface. There is no physical change in the state -- indeed, QFT is most naturally described in the Heisenberg picture -- but rather a shift in which state is appropriate for making predictions, given the information obtained from a measurement.

\section{Measurement and causality}\label{sec:impossible}

Another set of problems concerns the interaction between measurement and causality. This issue was raised in particular by Sorkin~\cite{sorkin1993impossible}, who attempted to generalise quantum mechanical ideal measurements to QFT, in particular retaining the projection postulate as the rule for updating states following measurement. Sorkin argued that this attempt fails because the measurement of apparently typical observables seems to lead to superluminal transfer of information, in defiance of causality. Therefore, many typical observables seem to require `impossible measurements' and one is left with the problem of determining, case by case, which observables can be measured and which cannot. We present a stripped down version of the scenario envisaged in~\cite{sorkin1993impossible}; see also~\cite{BorstenJubbKells:2021}. For this discussion, we work in a given Hilbert space and also describe states by density matrices.

Consider spacetime regions $O_1$, $O_2$ and $O_3$, so that the
causal future of $O_1$ intersects $O_2$, and the causal future of $O_2$ intersects $O_3$, but $O_1$ and $O_3$ have no direct causal connection. For this to occur it is necessary that the $O_i$ are extended regions, rather than points. We also assume that $O_1$ and $O_2$ can be separated by a Cauchy surface, as can $O_2$ and $O_3$. Suppose that Alice, Bob and Charlie can perform experiments in $O_1$, $O_2$ and $O_3$ respectively. 
Alice and Bob make nonselective measurements in their regions, represented for simplicity as operations on the state, using unitary operators $U_1$, $U_2$ localisable in $O_1$ and $O_2$ respectively.\footnote{Sorkin sidesteps questions as to where -- if at all -- the state update takes place by restricting to situations for which, in a certain sense, there is a well-defined causal order and each operation is concluded before the next begins.} Charlie's experiment measures an observable $C$ localisable in $O_3$. Alice chooses whether to perform her experiment, while Bob certainly performs his. So the two possible states in which Charlie will conduct his measurement are 
\begin{equation}
\rho_{BA}= U_2U_1\rho U_1^*U_2^*, \qquad\text{or}\qquad
\rho_B= U_2 \rho U_2^*,
\end{equation}
corresponding respectively to the cases in which Alice does or does not perform her experiment, where $\rho$ is the initial density matrix. Rotating traces, the corresponding expectation values for Charlie's measurement are 
\begin{equation}
    \langle C\rangle_{BA} = \Tr \rho U_1^*U_2^* C U_2 U_1,
    \qquad\text{or}\qquad
    \langle C\rangle_{B} = \Tr \rho U_2^* C U_2.
\end{equation}
Because $O_1$ and $O_3$ are causally disjoint, the operators
$U_1$ and $C$ commute. If $U_2$ happens to commute with either of $C$ or $U_1$ -- or more generally, if $U_1$ commutes with $U_2^*CU_2$ -- then $\langle C\rangle_{BA} =  \langle C\rangle_{B}$. However, because $O_2$ is not causally disjoint from either $O_1$ or $O_3$, there is no reason to assume that these commutators vanish,
and concrete examples show that $\langle C\rangle_{BA}$ and $\langle C\rangle_{B}$ differ in general as we will describe below, based on~\cite{Jubb:2022}.   
	Indeed, the typical expectation would be that the product $U_2^*CU_2$ cannot be localised in a smaller region than the causal hull of $O_2\cup O_3$.\footnote{An exception to this occurs in free field theories if $U_2$ is a Weyl operator (with arbitrary localisation) and $C$ is the hermitian part of a Weyl operator, in which case $U_2^*CU_2$ can be localised in $O_3$ and does not commute with typical unitaries localised in $O_1$.  See~\cite{Jubb:2022} for more on this example.}
In cases where $\langle C\rangle_{BA} \neq  \langle C\rangle_{B}$, Charlie can, in principle, determine whether Alice had performed her measurement, despite the fact that they are not in causal contact.

A concrete example can be given based on the theory of free scalar fields~\cite{Jubb:2022} -- see~\cite{MuchVerch:2023} for examples formulated under general conditions. Let $U_1=e^{i\phi(f)}$ and $U_2=e^{i\phi(g)^2}$ where $f$ and $g$ are real-valued test functions compactly supported in $O_1$ and $O_2$ respectively, and the notation $\phi(f)$ indicates a smeared field -- we assume a sufficiently regular Hilbert space representation so that these are self-adjoint. Also let $h$ be a real-valued test function compactly supported in $O_3$. A calculation using the Baker--Campbell--Hausdorff formula and the covariant commutation relations $[\phi(f_1),\phi(f_2)]=iE(f_1,f_2)\II$, where $E$ is the $c$-number commutator distribution, gives
\begin{equation}\label{eq:adU2star}
	U_2^* e^{i\phi(h)} U_2 = e^{i\phi(h+2E(g,h)g)}
\end{equation}
and, on using the Weyl relations $e^{i\phi(f_1)}e^{i\phi(f_2)} = e^{-iE(f_1,f_2)/2}e^{i\phi(f_1+f_2)}$, 
\begin{equation}
	U_1^*	U_2^* e^{i\phi(h)} U_2 U_1 =  e^{2iE(g,h)E(f,g)} U_2^* e^{i\phi(h)} U_2 .
\end{equation}
Replacing $h$ by $th$ and then differentiating in $t$ at $t=0$, one finds 
\begin{equation}
		U_1^*	U_2^* \phi(h) U_2 U_1 = U_2^* \phi(h) U_2 + 2E(g,h) E(f,g)\II. 
\end{equation}	
To arrange that $\langle C\rangle_{BA} \neq  \langle C\rangle_{B}$ for $C=\phi(h)$, it is now
enough to find $f$, $g$ and $h$ compactly supported in $O_1$, $O_2$ and $O_3$ respectively, so
that $E(g,h)$ and $E(f,g)$ are both nonzero. If $f$ and $h$ are found so that the supports
of $E^\text{ret}f$ and $E^{\text{adv}}h$ both intersect $O_2$, where $E^\text{adv/ret}$ are the advanced and retarded Green functions, then the required $g$ certainly exists, as can be seen on recalling that
$E(f_1,f_2)=\int f_1 (E^{\text{adv}}f_2-E^{\text{ret}}f_2) \,d\text{vol}$.

Based on the situation described above, Sorkin~\cite{sorkin1993impossible} concluded that ``the Hilbert space formulation of quantum field theory [is left] with no definite measurement theory''. (For similar considerations, see also~\cite{BeckmanGottesmanKitaevPreskill:2002}.) We will later describe an account of measurement for QFT with state update rules that are fully consistent with causality~\cite{FewVer_QFLM:2018}. In particular, the framework resolves the problem of impossible measurements~\cite{BostelmannFewsterRuep:2020,FewsterJubbRuep:2022}, 
not only in the three-party situation described above, but for much more general collections of parties that perform selective or nonselective measurement. Consequently, the message to be drawn from~\cite{sorkin1993impossible} is that the fact that a unitary operator is localisable in some region $O$ does not imply that it induces an operation that can be physically performed within $O$. This should not be a surprise: for instance, the Lagrangians that describe  \emph{local} fields with \emph{local} interactions constitute a very specific (and small) subset of all possible Lagrangian field theories. Similarly, the update rules relevant in QFT turn out to be of a special type that cannot violate causality because, in the notation above, $U_1$ always commutes with $U_2^*CU_2$.

Although Sorkin's example was formulated in QFT, it has been noted that exactly analogous effects can be found in \emph{classical} field theory~\cite{MuchVerch:2023}. Again, the problem arises from the use of `passive operations' (e.g., local rotations of Cauchy data) that are localisable in some region, but which cannot be physically performed there. One could therefore conclude that the main significance of Sorkin's example is to point out that the notion of local operation, both in classical and quantum theories, must be refined to incorporate an understanding of whether they can be actively performed by physical interactions.

Finally, the issue of impossible measurements and its resolution has attracted recent attention from philosophers of physics~\cite{PapageorgiouFraser:2023b}.

\section{Local and covariant measurement schemes and state updates}

We now present a recently developed framework for describing
measurement in quantum field theory that addresses the problems set out in the foregoing sections while being both general and mathematically rigorous. References for this section include the original paper~\cite{FewVer_QFLM:2018} and the further developments given in~\cite{BostelmannFewsterRuep:2020, FewsterJubbRuep:2022} and~\cite{Ruep_thesis:2023}. A short account in~\cite{Few_Regensburg:2015} provides a summary of~\cite{FewVer_QFLM:2018}. 
The main idea of this framework is to analyse the measurement of one QFT (the system) by another (the probe) in terms of an interaction between these theories that is localised within a compact spacetime region, and to derive as much as possible from basic assumptions. It draws on ideas taken from quantum measurement theory~\cite{Busch_etal:quantum_measurement}, but implements them for quantum fields in (possibly curved) spacetime. The aim is not to solve the measurement problem in full, but rather to understand the individual steps in the measurement chain. 

It is first necessary to be more specific about the mathematical structure of quantum field theory. We will use algebraic quantum field theory (AQFT) \cite{Haag:book, AdvAQFT,FewsterRejzner_AQFT:2019,EncycMP_AQFT_BuchholzFredenhagen:2025,EncycMP_QFTCST_Kay:2025} which is well-suited to 
conceptual questions and naturally adapts to curved spacetimes in a fully covariant way~\cite{BrFrVe03, FewVerch_aqftincst:2015}.
In this contribution we cannot define all technical concepts fully, and refer the reader to the literature cited.

Let $M$ be a globally hyperbolic spacetime. A subset $O$ of $M$ is \emph{causally convex} if it contains all causal curves in $M$ whose endpoints lie in $O$; we describe any open causally convex subset as a \emph{region}, so $M$ itself is also a region. In AQFT, a quantum field theory $\Af$ on $M$ consists of an assignment of a \emph{local algebra} $\Af(M;O)$ to each region $O\subset M$, subject to a number of conditions. Each local algebra is a unital $*$-algebra\footnote{One might additionally demand that they are $C^*$- or von Neumann algebras but we shall not do this here.} and the principal conditions that will concern us here are the following. 
\begin{description}
    \item[Isotony] If $O_1\subset O_2$ then $\Af(M;O_1)\subset \Af(M;O_2)$ and the algebras share a common unit; in particular, $\Af(M;O)\subset \Af(M):=\Af(M;M)$ for every region $O$, and all $\Af(M;O)$'s share a common unit.
    \item[Time slice] If $O_1$ contains a Cauchy surface for $O_2$ then $\Af(M;O_2)\subset \Af(M;O_1)$; in particular, if $O$ contains a Cauchy surface of $M$ then $\Af(M;O)=\Af(M)$.
    \item[Einstein causality] Local algebras of causally disjoint regions commute elementwise.
    \item[Haag property] If $K$ is compact,
    $O$ is any connected region containing $K$, and
    $A\in\Af(M)$ commutes with $\Af(M;N)$ for every region $N\subset K^\perp$, then $A\in \Af(M;O)$. Here $K^\perp=M\setminus (J^+(K)\cup J^-(K))$ denotes the causal complement of $K$.
\end{description} 

Any element of $\Af(M;O)$ is described as being \emph{localisable in $O$}, but it is clear from the conditions just stated that a typical element of $\Af(M)$ is localisable in many regions. 
Elements of $\Af(M;O)$ might include smeared fields, where the smearing test functions are compactly supported in $O$, or operators formed from such elements by algebraic manipulations. 
The states of the theory are linear functionals $\omega:\Af(M)\to\CC$ that are
positive (obeying $\omega(A^*A)\ge 0$ for all $A\in \Af(M)$) and normalised
(obeying $\omega(\II)=1$); if $A$ represents an observable, $\omega(A)$ is its expected measured value when the state is $\omega$. 

\subsection{Coupled systems and measurement schemes}\label{sec:measurementscheme}

Now suppose that we have QFTs $\Af$ and $\Bf$ defined on $M$ with local algebras $\Af(M;O)$ and $\Bf(M;O)$ respectively. We regard theory $\Af$ as the system of interest and $\Bf$ as a probe theory that will be used to measure observables of $\Af$. This can only be done if there is some interaction between the two theories, and so we posit that the actual measurement involves a \emph{coupled theory} $\Cf$, rather than the \emph{uncoupled combination} $\Uf=\Af\otimes\Bf$ of the system and probe, but so that $\Cf$ resembles $\Uf$ outside a compact set $K\subset M$, which will be called the \emph{coupling zone}. Technically, this is implemented by requiring that there are isomorphisms between 
$\Cf(M;O)$ and $\Uf(M;O)=\Af(M;O)\otimes \Bf(M;O)$ for each region $O$ outside the causal hull of $K$, and that these isomorphisms are compatible with the separate isotony inclusions of the theories $\Cf$ and $\Uf$. 
In particular, the  regions $M^\pm=M\setminus J^\mp(K)$ provide invariantly defined `in' and `out' regions, so there are isomorphisms $\chi^-:\Uf(M;M^-)\to \Cf(M;M^-)$ and $\chi^+:\Uf(M;M^+)\to \Cf(M;M^+)$ identifying the uncoupled and coupled theories at early and late times respectively. Since both $M^\pm$ contain Cauchy surfaces of $M$, these maps define isomorphisms $\gamma^\pm:\Uf(M)\to\Cf(M)$ by the time slice property. The comparison of these two maps precisely compares the dynamics of $\Uf$ and $\Cf$ and enters naturally to the description of measurement. In taking up the point {\bf (A)} above, it is important to appreciate the distinction of the coupled and uncoupled theories -- Landau and Peierls do not seem to make a clear distinction which leads 
to difficulties. 

Although the actual physics of the measurement process is described by the theory $\Cf$, it is much more convenient to describe it with reference to the theories $\Af$ and $\Bf$. A particular experiment might be described as follows: \emph{at early times the system and probe are independently prepared in states $\omega$ and $\sigma$ respectively; the system and probe are coupled in the bounded coupling zone $K$; subsequently -- at late times -- the probe is measured.}
The isomorphisms $\gamma^\pm$ allow one to make precise translations between such statements about the system and probe theory to statements about the coupled theory. Thus, the preparation of the system in state $\omega$ and probe in state $\sigma$ corresponds to a state $\omega\otimes \sigma$ of $\Uf(M)$. At early times, the appropriate analogue of this state for the coupled theory is 
the state $\undertilde{\omega}_\sigma$ of $\Cf(M)$ so that
\begin{equation}
    \undertilde{\omega}_\sigma(\gamma^- (A\otimes B)) = (\omega\otimes\sigma)(A\otimes B)= \omega(A)\sigma(B)
\end{equation}
for all $A\in\Af(M)$, $B\in\Bf(M)$. That is, $\undertilde{\omega}_\sigma=((\gamma^-)^{-1})^*(\omega\otimes\sigma)$. Meanwhile, if $B\in \Bf(M)$ is a probe observable, it corresponds to $\II\otimes B\in \Uf(M)$.  At late times, the appropriate analogue in $\Cf(M)$ is provided by $\widetilde{B}=\gamma^+(1\otimes B)$. The expected measurement value returned by the experiment is 
\begin{equation}
    \undertilde{\omega}_\sigma (\widetilde{B}) = (\omega\otimes\sigma)((\gamma^-)^{-1}\gamma^+ (\II\otimes B)) = (\omega\otimes\sigma)(\Theta  (\II\otimes B)) ,
\end{equation}
where the scattering map $\Theta= (\gamma^-)^{-1}\circ \gamma^+$ 
describes the dynamics of $\Cf$ relative to that of $\Uf$, as mentioned above.
An important point is that $\Theta$ acts trivially on $\Uf(M;M^+\cap M^-)$, i.e., on observables localisable in the causal complement $K^\perp$ of $K$, which reflects the locality of the interactions involved.
Of course, the aim of the experiment is to learn something about an observable of $\Af(M)$ in the state $\omega$. Defining a map $\eta_\sigma:\Uf(M)\to\Af(M)$ by linear extension of
\begin{equation}
    \eta_\sigma(A\otimes B) = \sigma(B)A
\end{equation}
for $A\in\Af(M)$, $B\in \Bf(M)$,
it is easily seen that
\begin{equation}
    \undertilde{\omega}_\sigma (\widetilde{B}) = \omega(\varepsilon_\sigma(B))
\end{equation}
holds for all $B\in\Bf(M)$ and all states $\omega$ and $\sigma$ of the system and probe, where 
\begin{equation}
\varepsilon_\sigma(B) = \eta_\sigma(\Theta (\II\otimes B)).
\end{equation}
In this way, our experiment becomes a \emph{measurement scheme} for a measurement of the \emph{induced system observable} $\varepsilon_\sigma(B)$ in state $\omega$. By construction, the measurement statistics of the actual measurement (of $\widetilde{B}$
in state $\undertilde{\omega}_\sigma$) and its fictitious counterpart ($\varepsilon_\sigma(B)$ in state $\omega$) have the same expectation; in general their other moments will differ. For example, the variance of the actual experimental results is greater than or equal to that of the fictitious observable. We summarise a few more properties of the induced observables~\cite{FewVer_QFLM:2018}.

\paragraph{Complete positivity} Given a preparation state $\sigma$, $\varepsilon_\sigma$ is a completely positive linear map from the probe algebra to the system algebra, also obeying $\varepsilon_\sigma(\II)=\II$, $\varepsilon_\sigma(B^*)=\varepsilon_\sigma(B)^*$. In particular, it maps self-adjoint elements to self-adjoint elements, and positive elements to positive elements, and these properties are stable under tensoring in a finite-dimensional matrix algebra. 

\vspace{-0.5\baselineskip}

\paragraph{Unsharpness}  
The map $\varepsilon_\sigma$ is not generally a homomorphism, which means that projections need not be mapped to projections; rather, 
a projection is typically mapped to an \emph{effect} -- that is, a positive operator bounded above by $\II$. An effect $E$ may be interpreted as describing a measurement with possible outcomes $0$ or $1$, with probability $\omega(E)$ of obtaining $1$ in state $\omega$. For any $k>0$, the $k$'th moment $\mu_k$ of a zero-one random variable is just equal to the probability of obtaining $1$, so the variance of the measurement given by $E$ is $\Var_\omega(E) = \mu_2-\mu_1^2 = \omega(E)- \omega(E)^2$. By comparison, the quantum mechanical dispersion of $E$ is $(\Delta_\omega E)^2 = \omega(E^2)-\omega(E)^2$, so the difference is $\Var_\omega(E)-(\Delta_\omega E)^2 = \omega(N(E))$, where $N(E)=E-E^2$ is called the noise operator of the effect. As $N(E)\ge 0$, one sees that the sharpest possible zero-one measurements are those with $E=E^2$, i.e., the projections, while a general effect is afflicted with additional noise and is, in this sense, unsharp.
See e.g., Section~9.3 in~\cite{Busch_etal:quantum_measurement} and~\cite{BuschHeinonenLahti:2004} for further discussion.

\vspace{-0.5\baselineskip}

\paragraph{Localisation} For all $B\in \Bf(M)$ the induced observable $\varepsilon_\sigma(B)$
    is localisable in any connected region containing $K$ (and therefore also its causal hull). The probe can only measure system observables localisable in regions where the system and probe are coupled.  
    For typical induced observables and couplings, one cannot expect a tighter localisation than the causal hull of $K$.

\paragraph{Implementation in models} 
The locally covariant measurement schemes can be implemented rigorously in some models~\cite{FewVer_QFLM:2018} and are amenable to explicit calculations. Furthermore, it has been shown in a model of a free scalar field as the system, that a dense set of observables admit measurement schemes of this type, and therefore all local observables admit asymptotic measurement schemes~\cite{FewsterJubbRuep:2022}.

\vspace{-0.5\baselineskip}

\paragraph{Causality} If $B\in\Bf(M;O)$ where $O\subset K^\perp$, then $\varepsilon_\sigma(B)=\sigma(B)\II$. Nothing can be learned from a probe observable that is out of causal contact with the location of the coupling.  
 
\subsection{States and update rules}

We now turn to the question of how a state should be updated after a measurement. Consider, first, a selective measurement: with the system 
in a certain state, an observable $A$ is measured and (some property of) the outcome is known, e.g., its value, or whether it belongs to some specified range of values. The theorist's problem is to predict, at least statistically, the outcomes of further measurements, using the knowledge of the initial state and the knowledge gleaned from measurements made so far. Thus it is a
matter of computing probabilities for future outcomes, conditioned on those already in hand and the original state. In circumstances where those conditional probabilities can be obtained as \emph{unconditional} probabilities in a modified state, it is reasonable to adopt that modified state as the updated state consequent upon the measurement. We emphasise that this should not be regarded as a physical transition, but rather as calculational device.

This viewpoint actually underlies the standard measurement rule of quantum mechanics~\cite{vonNeumann:MathFoundQM}. Suppose that a measurement is made of an observable $A$ with nondegenerate eigenvalues, resulting in outcome $a$. Were it known  that the measurement would return the same result with certainty, if it were repeated immediately, then the unique state that correctly predicts future outcomes would be the eigenvector corresponding to eigenvalue $a$. However, many measurements are not of this type, and whether any are perfectly repeatable is open to question. For these reasons, Davies and Lewis~\cite{DaviesLewis:1970} dropped the repeatability hypothesis and introduced the concept of an \emph{instrument} that produces more general state updates consequent upon the measurement outcomes. 

In the framework of Sec.~\ref{sec:measurementscheme}, let us suppose that $A$ and $B$ are effects of the system and probe. The probability that $A$ is observed (returns the value $1$) given that $B$ is observed is computed as 
\begin{equation}\label{eq:condprob}
    \Prob_\sigma(A|B;\omega) = \frac{\Prob_\sigma(A\&B;\omega)}{\Prob_\sigma(B;\omega)} 
     = \frac{(\If_\sigma(B)(\omega))(A)}{(\If_\sigma(B)(\omega))(\II)}
\end{equation}
where
\begin{equation}
    \If_\sigma(B)(\omega):A\mapsto (\omega\otimes\sigma)(\Theta(A\otimes B)).
\end{equation}
is a positive linear functional whenever $\omega$ is.  
We now summarise the properties of the map $\If_\sigma$, referring to~\cite{FewVer_QFLM:2018} for proofs and more detail, after first recalling some terminology from the literature.

\paragraph{Channels} If $\mathcal{A}$ and $\mathcal{B}$ are two unital $*$-algebras, then a linear map $T : \mathcal{A} \to \mathcal{B}$ which 
is completely positive and satisfies $T({\bf 1}_\mathcal{A})\le {\bf 1}_{\mathcal{B}}$ is called a {\it channel}; in particular, channels map effects of $\mathcal{A}$ to effects of $\mathcal{B}$. A channel is 
called {\it non-selective} (or {\it probability-preserving}, or {\it 
unital}) if $T$ maps the unit element of $\mathcal{A}$ to the unit element of $\mathcal{B}$.

\paragraph{Operations} Considering again two unital $*$-algebras $\mathcal{A}$ and $\mathcal{B}$, we denote by $\mathcal{S}(\mathcal{A})$ the {\it state space} of $\mathcal{A}$, i.e.\ the set of all linear maps 
$\omega: \mathcal{A} \to \mathbb{C}$ which fulfill $\omega(A^*A) \ge 0$ for all $A \in \mathcal{A}$, $\omega({\bf 1}_{\mathcal{A}}) = 1$ and which 
are continuous when $\mathcal{A}$ is endowed with a topology. (Continuity follows from the previous requirements on $\omega$ e.g.\ if $\mathcal{A}$ is a $C^*$-algebra.) We also introduce the {\it sub-state space} $\mathcal{S}^{\vee}(\mathcal{A})$ which consists of functionals $\omega$
on $\mathcal{A}$ fulfilling the previous properties of a state except for allowing the more general condition $\omega({\bf 1}_{\mathcal{A}}) \le 1$. Using the analogous terminology for $\mathcal{B}$ in place of $\mathcal{A}$, an {\it operation} is a convex map $\tau : \mathcal{S}^{\vee}(\mathcal{B}) \to \mathcal{S}^{\vee}(\mathcal{A})$. 
An operation $\tau$ is called {\it non-selective} if it induces a map 
$\tau: \mathcal{S}(\mathcal{B}) \to \mathcal{S}(\mathcal{A})$. Obviously, 
if $T : \mathcal{A} \to \mathcal{B}$ is a (non-selective) channel, then
its dual $T^* : \mathcal{S}^{\vee}(\mathcal{B}) \to \mathcal{S}^\vee(\mathcal{A})$ given by $(T^*(\omega))(A) = \omega(T(A))$ is a (non-selective) operation. This is the typical way in which operations arise in the von Neumann algebraic context. 

\paragraph{Pre-instruments} In \cite{FewVer_QFLM:2018}, we have introduced the concept of a {\it pre-instrument} as a  map $\mathcal{J}$ from the 
effects in a unital $*$-algebra $\mathcal{B}$ to the operations of another 
unital $*$-algebra $\mathcal{A}$. In other words, for every effect 
$B \in \mathcal{B}$, $\mathcal{J}(B) : \mathcal{S}^{\vee}(\mathcal{A}) \to \mathcal{S}^\vee(\mathcal{A})$ is an operation. According to Davies and Lewis \cite{DaviesLewis:1970}, an {\it instrument} $\mathcal{I}$ is a map
defined on a $\sigma$-algebra $\Xc$ of some total space $\Omega$, with
values in the operations of a unital $*$-algebra $\mathcal{A}$, having additivity properties like a measure.  If we consider 
any measure $\Es$ on $\Xc$ valued in the effects of $\mathcal{B}$ (i.e.\ a positive-operator-valued measure, or POVM) together 
with a pre-instrument $\mathcal{J}$ as described before, we require -- as 
an additional property of pre-instruments -- that
$\mathcal{I} = \mathcal{J} \circ \Es$ is an instrument in the Davies--Lewis sense. 

\paragraph{State updates}  Returning to the map $\If_\sigma$ defined above,  if $B$ is an effect,  $\If_\sigma(B)$ is dual to the channel $A\mapsto \eta_\sigma\Theta (A\otimes B)$ of $\Af(M)$ and therefore defines an operation 
 on $\mathcal{S}^\vee(\Af(M))$. If $\Es$ is an effect-valued measure as before, then for any
$X \in \Xc$, 
$\omega_{\Es(X),\sigma} = \If_\sigma(\Es(X))(\omega)$ in $\mathcal{S}^\vee(\Af(M))$ gives
the (in general, selectively) updated (sub-)state conditioned on the probe measurement results  
of $\Es$ lying in $X$.

\paragraph{Selective updates} Together with the observation just made, Eq.~\eqref{eq:condprob} shows that
\begin{equation}
    \omega_{B,\sigma} = \frac{\If_\sigma(B)(\omega)}{(\If_\sigma(B)(\omega))(\II)} 
\end{equation}
is a state of $\Af(M)$ with the property that
\begin{equation}
    \Prob_\sigma(A|B;\omega) = \omega_{B,\sigma}(A).
\end{equation}
In other words it correctly predicts outcome probabilities conditioned on the effect $B$ being observed successfully, and therefore deserves to be called a \emph{selectively updated state} conditioned on success of $B$. 
We mention that the notation for the updated state should, more precisely, be 
$\omega_{\Theta,B,\sigma}$ since the updated state clearly depends on the 
interaction between system and probe which is described by the scattering map $\Theta$. However, this is inconvenient, and so we drop the indication of $\Theta$
in what follows. 

\paragraph{Nonselective updates} If $\Omega$ is the total space of $\Es$, then $\Es(\Omega)=\II$, which is the trivially successful effect. That is, no selection on the measurement results occurs at all. The corresponding state $\omega_{\text{ns}}=\omega_{\Es(\Omega),\sigma}$ depends only on the scattering map $\Theta$ and the probe preparation state $\sigma$ as the formula
\begin{equation}
    \omega_{\text{ns}}(A) = (\omega\otimes\sigma) (\Theta A\otimes \II) 
\end{equation}
makes plain. Although no information from the measurement is used, the system and probe have still been coupled together in a particular way. In the effective `system only' description, we trace out the probe degrees of freedom in the `out' region, thus obtaining a modified system state. 
Again we mention that 
a more detailed notation might be $\omega^{\text{ns}}_{\Theta,\sigma}$ including
the dependence on the scattering map $\Theta$ and the initial probe state $\sigma$, but we shall not employ this notation. 
\\ \\
Two results illustrate the contrast between selective and nonselective updates. First, for all system observables $A$ localisable spacelike to the coupling zone $K$ one has $\omega_{\text{ns}}(A)=\omega(A)$. That is, nonselective updates do not affect the predictions of spacelike separated observers. Second,  for a selective update conditioned by successful measurement of probe effect $B$ that is localisable spacelike to $K$, it may be shown that $\omega_{B,\sigma}(A)=\omega(A)$ if and only if $A$ is uncorrelated with $\varepsilon_\sigma(B)$ in the system state $\omega$. Thus selective updates can affect predictions of spacelike separated observers, but only due to pre-existing correlations in the system state.

\subsection{Multiple probes}

The scheme sketched above provides a satisfactory account of one step in the measurement chain for a single measurement. This can be extended in two ways: first, to understand measurements in which there are several steps in a chain of measurement, and second to understand multiple probes. In fact these boil down to the same basic problem of coupling multiple probes to a system in such a way that they can be discussed individually or collectively. 
Consider a system theory and a finite collection of probe theories, each with its own coupling zone and scattering map. A \emph{causal order} on the set of coupling zones is a strict linear order $\triangleleft$ so that $K\triangleleft K'$ only if $J^-(K)$ and $J^+(K')$ do not intersect, or equivalently, that there is a Cauchy surface with $K$ to its past and $K'$ to its future. If the coupling zones admit at least one causal order then we say they are \emph{causally orderable}; it can happen that there are many causal orders on a given set of coupling zones, as can be seen in the example of two spacelike separated regions, for which either order is possible.  

Suppose that one has two probe theories $\Bf_1$ and $\Bf_2$, each of which can be combined with the system theory $\Af$ to produce coupled theories $\Cf_1$ and $\Cf_2$ with coupling zones $K_1$ and $K_2$. Following the procedure described above, the dynamics of these coupled systems can be compared with that of the uncoupled combinations resulting in scattering maps $\Theta_1$ and $\Theta_2$ on $\Af(M)\otimes\Bf_1(M)$ and $\Af(M)\otimes\Bf_2(M)$. On the other hand, the two probes could be combined as a `super probe' theory $\Bf_1\otimes \Bf_2$. We suppose that there is a coupled theory that describes the system coupled to both probes, with coupling zone $K_1\cup K_2$,  whose dynamics are captured by a scattering map $\Theta$ on $\Af(M)\otimes \Bf_1(M)\otimes \Bf_2(M)$. This assumption is physically reasonable when the coupling zones are well-separated, but would not necessarily hold if they overlap, for example.  

To analyse the combined measurement it is necessary to relate $\Theta$ to $\Theta_1$ and $\Theta_2$. 
In the situation where $K_1$ and $K_2$ are causally orderable, with $K_1\triangleleft K_2$, a natural assumption is that 
\begin{equation}
\Theta = \hat{\Theta}_1\circ \hat{\Theta}_2  
\end{equation}
where $\hat{\Theta}_1 = \Theta_1\otimes \id$ and $\hat{\Theta}_2 = \Theta_2\otimes_2\id$ lift the individual scattering maps to $\Af(M)\otimes \Bf_1(M)\otimes \Bf_2(M)$
(by convention, $(\alpha\otimes\beta)\otimes_2 \gamma := \alpha\otimes\gamma\otimes\beta$). Note that the later scattering map is executed first, and also that the above \emph{causal factorisation} is required for any admissible causal order on the coupling zones. In particular, if $K_1$ and $K_2$ are spacelike separated, then one has
$\Theta = \hat{\Theta}_1\circ \hat{\Theta}_2  = \Theta = \hat{\Theta}_2\circ \hat{\Theta}_1$. In our axiomatic development, causal factorisation appears as an additional assumption, but it is well-established in quantum field theory models~\cite{Rejzner_book,Duetsch_book} and, more recently, has been taken as a starting-point for a novel construction of interacting models~\cite{BuchholzFredenhagen:2020}.

With these preparations, the relation between the pre-instruments associated with the individual measurements, and the pre-instrument associated with a combined measurement of the two probes can be given as follows
\begin{equation}
\If_{\sigma_2}(B_2)\circ \If_{\sigma_1}(B_1)= \If_{\sigma_1\otimes\sigma_2}(B_1\otimes B_2).
\end{equation}
Here, $\sigma_1$ and $\sigma_2$ are the two probe preparation states, 
$B_1\in \Bf_1(M)$, $B_2\in\Bf_2(M)$ are probe observables, and we are again assuming that $K_1\triangleleft K_2$. Whereas scattering maps are composed with the latest interaction first, the pre-instruments are combined from past to future.
A crucial observation is that if either causal order is admissible, then the pre-instruments may be composed in either order, with the same result.

\paragraph{Successive selective updates} A direct consequence of the above is that, for $K_1\triangleleft K_2$, the state obtained by successive updates from successful measurements of  effect $B_1$, and then a successful measurement of effect $B_2$, coincides with the update for a successful measurement of $B_1\otimes B_2$, i.e.,
\begin{equation}
    (\omega_{B_1,\sigma_1})_{B_2,\sigma_2}  = \omega_{B_1\otimes B_2,\sigma_1\otimes\sigma_2}.
\end{equation}
The updates may be performed in either order if $K_1$ and $K_2$ are spacelike separated.

\paragraph{Multiple probes} When $N>2$ probes are considered, we assume that the probes may be combined in arbitrary clusters that may themselves be combined, subject always to causal orderability, and with $N$-fold combinations reducing to the bi-partite case. More details are given in~\cite{BostelmannFewsterRuep:2020}.
The conclusion is that the selective state updates provide correct conditional expectation values conditioned on the success of separate probe measurements, 
\begin{equation}
\mathbb{E}(B_{N+1}| B_1 \& B_2 \& \cdots\& B_N;\omega) = 
\mathbb{E}(B_{N+1}; ((\omega_{B_1})_{B_2})_{\cdots B_N})
\end{equation} 
if probe effects $B_1,\ldots B_{N+1}$ are measured by causally orderable probes with coupling zones $K_1\triangleleft  \cdots \triangleleft K_{N+1}$; and that the equation is true for all valid causal orderings of the coupling zones. (We have suppressed the probe preparation states $\sigma_j$ in the above equation.)

Related results hold for nonselective state updates. For example, 
suppose that there are $M+N+1$ causally orderable coupling zones, with
\begin{equation}
K_{1}\triangleleft\cdots\triangleleft K_{M} \triangleleft  K_{M+1}
\triangleleft K_{{M+2}}
\triangleleft\cdots \triangleleft K_{{M+N+1}}
\end{equation}
and corresponding probe effects $B_1,\ldots,B_M$ and $B_{M+2},\ldots B_{M+N+1}$ are measured without selection, then the expectation of any system observable $B_{M+1}$ induced by a measurement of an observable of the $M+1$'st probe is given by
\begin{equation}
\mathbb{E}(B_{M+1};\omega) = \mathbb{E}(B_{M+1}; 
((\omega_{\text{ns}}^{(1)})_{\text{ns}}^{(2)})_{\cdots\text{ns}}^{\cdots (M)}
),
\end{equation}
where $\omega\mapsto \omega^{(j)}_{\text{ns}}$ indicates the nonselective update induced by the coupling in $K_j$. Thus, the expectation value depends on the  measurements ordered to the past of $K_{M+1}$, but not on those ordered to its future. Once again, we emphasise that these physically natural results are independent of the specific casual order used, other things being equal.

\paragraph{Absence of impossible measurements} Using the analysis of multiple probes, together with refined information on the locality properties of   scattering maps, one can address the question of impossible measurements (see Section~\ref{sec:impossible}) in a precise way. Recall that the scenario 
involves experimental regions $O_1$, $O_2$ and $O_3$ with $O_1$ and $O_3$ causally disjoint but $O_2$ overlapping the causal future of $O_1$ and the causal past of $O_3$.  Meanwhile there are Cauchy surfaces separating $O_1$ from $O_2$, and $O_2$ from $O_3$. Alice, the experimenter in region $O_1$, chooses whether or not to make a nonselective measurement with coupling zone $K_1\subset O_1$, while Bob certainly makes a nonselective measurement in region $O_2$. One may now prove~\cite{BostelmannFewsterRuep:2020} that Charlie, experimenting in region $O_3$, cannot determine whether or not Alice had made her measurement. There are no impossible measurements in this framework, and this result extends to situations involving arbitrarily many observers with causally orderable coupling zones.
Since every local observable, at least in simple models, is (asymptotically) measureable in this way, it seems that the problem raised by Sorkin~\cite{sorkin1993impossible} has been resolved.  

As already mentioned, the message to be drawn is that algebra elements localisable in a region $O$ need not represent local operations that can be undertaken there. On the other hand, hermitian algebra elements have a good interpretation as observables. This casts an interesting light on the interpretation of local algebra elements in AQFT.

\section{Accelerated detectors}\label{AccDetc}

We now address the large body of literature concerning a detector model that has gained considerable prominence since it was introduced by Unruh~\cite{Unruh:1976} and subsequently refined by many other authors. For a review, see~\cite{CrispinoHiguchiMatsas:2008}.

The common feature of these models is that a quantum field is coupled to a quantum mechanical probe along (or in the vicinity of) a timelike curve in spacetime. The famous Unruh effect is that an eternally coupled probe following a uniformly linearly accelerated trajectory in Minkowski spacetime becomes excited into a thermal state when the quantum field is in its Poincar\'e-invariant vacuum state. While the majority of the literature proceeds at the level of first order perturbation theory, the same overall result has been obtained with full rigour by de Bi\`evre and Merkli~\cite{deBievreMerkli:2006}, for a suitable probe interaction.

From the perspective of mathematical physics, the traditional description of the Unruh detector presents several problems. Consider a classical massless scalar field $\phi$ (the system) coupled along an inertial curve $(t,\boldsymbol{0})$ in Minkowski spacetime to a classical harmonic oscillator with displacement $\psi$ (the probe) with combined equations of motion 
	\begin{align}
	(\Box \phi)(t,\boldsymbol{x}) + \lambda \delta(\boldsymbol{x})\psi(t) &=0\\
	\ddot{\psi}(t) + \omega^2 \psi(t) +\lambda \phi(t,\boldsymbol{0})&=0,
	\end{align} 
where $\lambda$ controls the strength of the coupling. The retarded solution to the first equation is
	\begin{equation}
	\phi(t,\boldsymbol{x}) =-\lambda \int dt'\,d^3\xb'\, \frac{\delta(t-t'-\|\xb-\xb'\|)}{4\pi\|\xb-\xb'\|} \delta(\xb')\psi(t') = -\lambda
	\frac{\psi(t-\|\xb\|)}{4\pi \|\xb\|},
	\end{equation}
from which it is clear that $\phi(t,\boldsymbol{0})$ is ill-defined if $\psi(t)\neq 0$, so
the only nonsingular solutions are those in which $\psi(t)\equiv 0 \equiv\phi(t,\boldsymbol{0})$; in other words, those in which there is no interaction between system and probe.  With this in mind,  it is now unclear what well-defined combined system is approximated by the perturbative quantum treatment.

Various modifications have been discussed. One is to invoke a short-distance cutoff~\cite{HuLin:2006}, while another is to smear out the coupling between field and oscillator (or other probe) around a timelike curve~\cite{GroveOttewill:1983,Takagi:1986,Schlict:2004}.
Indeed the latter is essential to the rigorous mathematical treatment in~\cite{deBievreMerkli:2006}. The smeared coupling mollifies the solutions and produces a well-posed classical dynamics. However, it comes at a cost, because spacelike separated points are coupled together through the interaction. Thus the dynamics does not respect the causal structure of the background spacetime, although the deviations become smaller as the finite extent of the detector is shrunk.
Nonetheless, the fact that the dynamics is acausal renders such models less suitable for discussion of causality in quantum field theoretic measurement, although the framework is sufficiently general to allow this~\cite{PoloGomezGarayMartinMartinez:2022}. 
Note that there can be important differences in the behaviour of probes described by quantum fields and those described by quantum mechanical systems~\cite{Ruep:2021}.

An important question is: what observables of the quantum field are measured by such detector models, and where are they localised? In fact this was the original motivation for the framework of~\cite{FewVer_QFLM:2018} described above. If one considers a single mode in a probe quantum field instead of a harmonic oscillator, one may with some approximation, retrieve the usual first-order perturbative treatment~\cite{FewVer_QFLM:2018}. This suggests that the localisation of induced observables should follow similar rules to those explained in Section~\ref{sec:measurementscheme}. One is thereby led to conclude that observables induced by an eternally coupled detector following a linearly accelerated trajectory have no smaller localisation region than the full Rindler wedge generated by the trajectory. 
 
In particular, the exact thermalisation of the detector requires an eternal coupling between system and probe, and therefore tests system degrees of freedom from a whole wedge, rather than just in the immediate vicinity of the detector trajectory. This relates to the fact that the Poincar\'e-invariant vacuum state is a KMS state at the Unruh temperature with respect to the boost `time' for a wedge region~\cite{Fulling73,BisognanoWichmann:1975,Sewell}.

A much-discussed theme in the literature is the sense (if any) in which accelerated observers locally perceive the vacuum as a thermal bath. It has been pointed out that this is a highly problematic concept even from thermodynamic considerations~\cite{BellHughesLeinaas:1985,BucVer_macroscopic:2015,BuchholzVerch:2016}. The disussion above adds a further aspect to this, namely that the induced detector observables are not localisable close to the detector trajectory. To be clear: the detector will thermalise asymptotically~\cite{deBievreMerkli:2006}; the issue is whether and how this is reflected in local measurements.  

Actually, some measures of thermality in the sense of detailed balance can be approximately attained after sufficient coupling time~\cite{FewsterJuarezAubryLouko:2016} and therefore correspond to measurements in finite, though growing, spacetime regions. Incorporating the asymptotic limit into a framework such as~\cite{FewVer_QFLM:2018} is an important open question for which some steps are in hand~\cite{FewsterJubbRuep:2022}.

\section{Some related approaches} \label{RelAppr}

To conclude, we briefly touch on some related literature on measurement in algebraic quantum field theory. 

A long time ago, Hellwig and Kraus~\cite{HellwigKraus:1969,HellwigKraus:1970} also considered measurement in terms of an interaction between a quantum field and an apparatus (an external quantum system -- a probe, in our terminology), whose interaction with the system is described by a scattering operator. They develop the concept of a local operation in that setting. More recently, Doplicher \cite{DoplicherQFM} has considered measurement from the perspective of a quantum field theory that provides both the system and the probe. This is physically well-motivated, by the viewpoint that both the system and probe in an actual laboratory should in principle be describable within the standard model of elementary particle physics. From this perspective, our introduction of a separate probe theory is an idealisation. What is less clear in the approaches of both Hellwig--Kraus and Doplicher is how the more fine-grained localisation of induced observables and corresponding instruments arises, because the interaction is not modelled via a local quantum field theory.

Although the algebraic notion of a state is linked to the Hilbert space state vector concept through the GNS representation, this correspondence does not directly allow for a natural notion of superposition of states or transition probabilities in the sense of Wigner. This problem is acute when the type of the local algebras is properly infinite as is the case in local quantum field theory~\cite{Fredenhagen:1985}. Buchholz and St{\o}rmer~\cite{BucSto:2014} have addressed this problem in contexts where the local algebras are approximated by `funnels' of type $\mathrm{I}_\infty$ von Neumann algebras, and have developed natural notions of both superposition and transition probability between states. 
 
Finally, Okamura and Ozawa \cite{OkamuraOzawa,Okamura:2021} consider refined concepts of local instruments in algebraic quantum field theory. They show that local instruments of a vacuum representation induce local instruments in 
localisable superselection sectors~\cite{Haag:book,EncycMP_AQFT_BuchholzFredenhagen:2025,EncycMP_SymmetriesSuperselection_Rehren:2025} under certain conditions.  
%

\end{document}